# "Visible" 5d orbital states in a pleochroic oxychloride

Daigorou Hirai[*,†], Takeshi Yajima[†], Daisuke Nishio-Hamane[†], Changsu Kim[†], Hidefumi Akiyama[†], Mitsuaki Kawamura[†], Takahiro Misawa[†], Nobuyuki Abe[‡], Taka-hisa Arima[‡], and Zenji Hiroi[*,†]

[†]Institute for Solid State Physics, University of Tokyo, Kashiwa, Chiba 277-8581, Japan

[‡]Department of Advanced materials science, University of Tokyo, Kashiwa, Chiba 277-8561, Japan

**ABSTRACT:** Transition metal compounds sometimes exhibit beautiful colors. We report here on a new oxychloride $Ca_3ReO_5Cl_2$ which shows unusually distinct pleochroism; that is, the material exhibits different colors depending on viewing directions. This pleochroism is a consequence of the fact that a complex crystal field splitting of the $5d$ orbitals of the $Re^{6+}$ ion in a square-pyramidal coordination of low-symmetry occurs accidentally in the energy range of the visible light spectrum. Since the relevant $d$-$d$ transitions possess characteristic polarization dependences according to the optical selection rule, the orbital states are "visible" in $Ca_3ReO_5Cl_2$.

## INTRODUCTION

Color is a fundamental property of a material. Beautiful colors of gemstones have attracted people for thousands of years. It is also important for materials science, as it carries information on quantum mechanical states that govern the basic properties of materials. The color of a transition metal compound is mostly associated with electronic transitions between well-defined $d$ levels ($d$-$d$ transitions), which are induced by the absorption of light with corresponding energies ranging from 1.5 to 3.5 eV in the visible light spectrum. The $d$ levels of an isolated transition metal ion are fivefold degenerate, while the degeneracy is partly or completely lifted by an electrostatic field from the surrounding ligands to induce a crystal field splitting (CFS). Thus, the color can probe the hierarchy of $d$ levels with appropriate CFSs, which is crucial for understanding the chemical bonding and physical properties of materials. For example, ruby's red is important for the laser generation and is known to be caused by the absorption of yellow-green light, which is complementary to red, by a $d$-$d$ transition between split $3d$ states of a $Cr^{3+}$ ion substituted for Al in $Al_2O_3$[1–3]. $Cr^{3+}$ ions are also responsible for the green color of emerald, having a smaller CFS with the absorption of red light[4]. These beautiful colors of gemstones rely on the delicate choice of energy and strength of absorptions and are wonderful gifts from nature.

The color of transition metal compounds is also helpful in our daily life. For example, cobalt chloride is widely used as an indicator of water for desiccants: the cobalt ion is coordinated by four chlorine ions in $[CoCl_4]^{2-}$ when dry, while by six water molecules in $[Co(H_2O)_6]^{2+}$ when hydrated. This change in the coordination of the $Co^{2+}$ ion makes the CFS larger and causes a color change from blue to pink with hydration[5].

A transition metal compound would not always exhibit the color even if the CFS lies in the energy range of visible lights. This is because the optical selection rule constrains possible electronic transitions according to the relationship between the parities of orbitals at the ground state and exited states with respect to the polarization of light (electronic field of light). Because the parity of a $d$ orbital is determined by the electro-static field of surrounding ligands, the coordination of a transition metal ion is crucial for the appearance of color. In many transition metal oxides, the transition metal ion is typically surrounded by six oxide ions in an octahedral coordination. In a centrosymmetric regular octahedron, $d$-$d$ transitions are strictly forbidden because the parities of all the $d$ orbitals are identically "even", which is known as the Laporte rule[6]. Transitions are allowed when the octahedron distorts either statically or dynamically. Thus, transition metal compounds with non-centrosymmetric coordination polyhedra can have strong absorptions of light. An excellent example is the recently discovered vibrant blue pigment $YIn_{1-x}Mn_xO_3$[7]. In this compound, the Mn ion has a rare trigonal bipyramidal coordination without inversion symmetry, and symmetry-allowed $d$-$d$ transitions result in the intense absorption of visible lights. New inorganic pigments with a variety of colors have been found by focusing on this unique trigonal bipyramidal coordination[8–10].

One of the interesting optical phenomena that a crystal exhibits is pleochroism. In principle, all crystals except ones with the cubic symmetry can have different colors depending on the direction of observation or the polarization of light with respect to the crystallographic axes. However, pleochroism is weak in most crystals. Strong color changes are exceptionally fond in some minerals like alexandrite[11], tourmaline, and cordierite (iolite)[12–14]; the pleochroism is used to identify these minerals or to make them into accessories by cutting along appropriate directions. The reason why pleochroic crystals with intense and beautiful color changes are limited in nature may be that an accidental balance is required on the energy, intensity, and anisotropy of light absorptions allowed by crystal symmetry.

In this study, we report on a new $5d$ transition metal oxychloride $Ca_3ReO_5Cl_2$ which shows a strong pleochroism. $5d$ compounds have been less investigated compared to $3d$ compounds. However, they attract great attention in recent years because they show unique physical properties based on the strong spin-orbit coupling inherent to heavy $5d$ elements. In the course of materials exploration in $5d$ compounds, we have found that a crystal of $Ca_3ReO_5Cl_2$ exhibits three different vivid colors of green, red, and yellow for lights polarized along the crystallographic *a*, *b*, and *c* axes, respectively. We show that this unique optical property stems from the optical selection rule of electronic transitions between the split 5*d* states of the Re$^{6+}$ ion in a specific square-pyramidal coordination. Ca$_3$ReO$_5$Cl$_2$ provides us with a rare example in which quantum mechanical orbital states are "seen" as beautiful colors.

### EXPERIMENTAL SECTION

Single crystals were grown by a flux method. CaO, ReO$_3$, and CaCl$_2$ in a molar ratio of 3:1:4.8 were mixed in an agate mortar in an argon-filled glove box, and the mixture was put in a gold tube and sealed in an evacuated quartz ampoule. The ampoule was heated at 1000 °C for 24 hours and then slowly cooled to 800 °C in a cooling rate of 1 °C/hr. Several crystals of approximately 5 mm$^3$ volume were obtained after excess CaCl$_2$ flux was washed away by distilled water. The crystal has a cleavage plane perpendicular to the *a* axis. The compound is practically stable, but decomposes in a week in air possibly by the effect of moisture.

Chemical analysis by means of energy dispersive x-ray spectroscopy were performed in a scanning electron microscope (JEOL JSM-5600) operated at 15 kV and 0.4 nA with beam diameter of 2 μm, and the ZAF method was used for data correction. Single crystal x-ray diffraction (XRD) measurements were conducted at room temperature using a R-AXIS RAPID IP diffractometer (Rigaku) with a monochromated Mo-*Kα* radiation. The structure was solved by direct methods and refined by full-matrix least-square methods on $|F^2|$ by using the SHELXL2013 software.

Absorption spectra were recorded in an Irtron IRT-30 Infrared Microscope (JASCO) in the wavenumber range between 0.4 and 3.2 eV. A halogen lamp (0.5 – 3.5 eV) was used as a light source, and a Glan-Taylor prism was used as a polarizer. Detectors used are PbS (photoconductive element) for the near-infrared region and a photomultiplier tube from the ultraviolet to visible region. A sample was placed on a glass slide and transmission measurement was performed with an aperture of 150 × 150 μm$^2$.

X-ray absorption spectroscopy (XAS) experiments were performed around Re $L_{III}$ edge on BL19LXU at SPring-8, Japan. A single crystal was irradiated by linearly polarized x-ray beam at room temperature, and the intensity of Re L luminescence was measured by sweeping the incident x-ray energy from 10.48 to 10.62 keV with an increment of 1 eV.

First-principles calculations were performed based on the density functional theory (DFT), using a program package Quantum ESPRESSO[15] which employs plane-waves and psudopotentials to describe the Kohn-Sham orbitals and the crystalline potential, respectively. The plane-wave cutoff for a wavefunction was set to 60 Ry. Calculations were performed with a GGA-PBE functional[16] using ultrasoft pseudopotentials[17]. We set *k*-point grids of Brillouin-zone integrations for the charge density and the partial density of states to 5×10×5 and 10×20×10, respectively. Wannier functions were obtained by using a program package Wannier90[18] which computes the maximally localized Wannier orbital.

### RESULTS AND DISCUSSION

We have obtained two kinds of crystals with green and brown/orange (depending on the thickness) colors, as shown in Fig. 1(a). First we thought that two kinds of compounds were produced, but we have noticed that they are identical to each other, because a green crystal becomes brown when it is rotated by 90°, (a supplemental movie). Thus, the crystal exhibits distinct pleochroism.

The powder XRD pattern from crushed crystals does not match any pattern for known compounds in the database. Chemical analysis found the molar ratio of Ca, Re, and Cl to be Ca:Re:Cl = 3:1.12:1.98. Oxygen was also detected in the measurement, but the amount could not be determined quantitatively because of poor sensitivity for light elements.

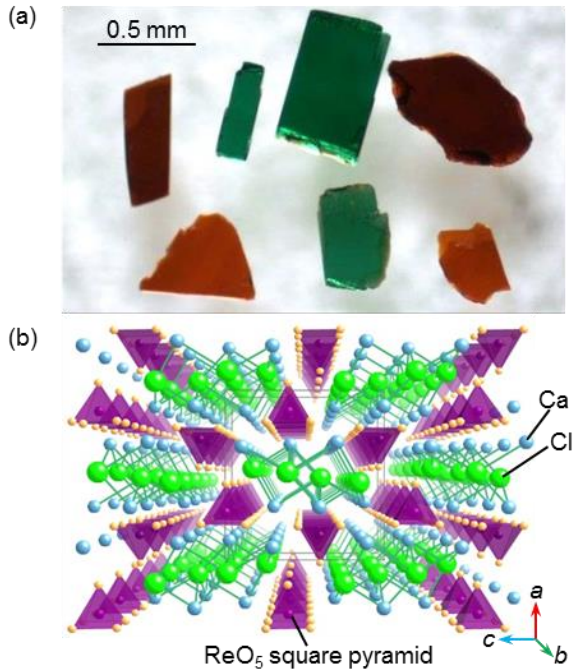

**Figure 1.** (a) Photograph of single crystals of Ca$_3$ReO$_5$Cl$_2$. They show two kinds of colors, green and brown/orange, depending on the viewing direction. (b) Crystal structure of Ca$_3$ReO$_5$Cl$_2$ viewed perspectively along the [010] direction. It crystallizes in an orthorhombic structure with the space group *Pnma*. The ReO$_5$ units are shown by purple square pyramids, and the calcium and chlorine atoms are represented by large blue and small green spheres, respectively. The ReO$_5$ square pyramids have no common vertices with each other to form the three-dimensional network.

It is revealed by single crystal XRD measurements that this compound is isostructural to the known compound Ca$_3$WO$_5$Cl$_2$ crystallizing in an orthorhombic structure [space group *Pnma*; lattice constants: *a* = 11.820(2) Å, *b* = 5.587(1) Å, and *c* = 11.132(1) Å; the original lattice constants were given for the space group *Pnam*][19]. This suggests that a Re analogue, Ca$_3$ReO$_5$Cl$_2$ with lattice constants *a* = 11.8997(2) Å, *b* = 5.5661(1) Å, and *c* = 11.1212(2) Å has been produced. This is consistent with the result of chemical analysis within the standard experimental error. Moreover, the oxygen content must be 5 per formula unit, as no vacancy is observed at the O site in the structural refinement. Thus, the valence state of the Re ion is 6+ in the the electronic configurations of 5*d*$^1$.

A unique feature of this orthorhombic crystal structure is that the Re$^{6+}$ ion is surrounded by five O$^{2-}$ ions in a rare square-pyramidal coordination, as shown in Fig. 1(b). There is a $Cl^-$ ion on the opposite side of the apical oxygen of the $ReO_5$ square pyramid. However, the distance from the $Re^{6+}$ to the $Cl^-$ ion is 3.558(3) Å, which is much larger than the typical bond length between $Re^{6+}$ and $Cl^-$ of ~2.2 Å[20]. Thus, it is appropriate to consider the local coordination of $Re^{6+}$ as the $ReO_5$ square pyramid (SP) rather than the $ReO_5Cl$ octahedron. Note that the SP has a small trapezoidal distortion of the basal plane so as to lose one of the mirror planes (local symmetry at the Re site is .m.), which would be critical to consider the light absorption later. Since there is only one crystallographic site for Re, all the SPs are identical. The half of them point upward along the [100] direction and the other half point downward along the [$\bar{1}$00], which appear in a staggered manner along the $c$ axis. The $ReO_5$ units are isolated from each other without sharing their oxygens and are separated by $Ca_3Cl_2$ slabs running along the $b$ direction. Note that the arrangement of the $ReO_5$ units is three dimensional, though one dimensional arrays are noticed in the picture of Fig. 1 (b): the neighboring Re-Re distances are 5.5661(1) Å along the $b$ axis, while 5.5515(3) and 6.3989(3) Å along the other two directions.

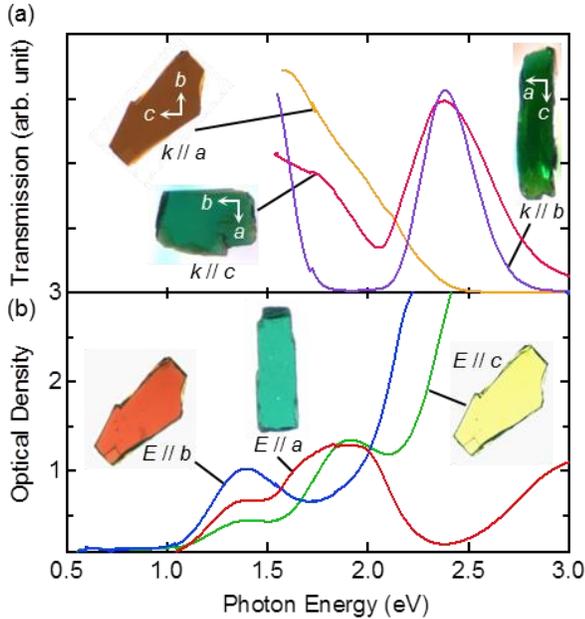

**Figure 2.** (a) Transmission spectra for unpolarized lights propagating along the $a$, $b$, and $c$ directions and the corresponding images of crystals. (b) Optical densities for incident lights polarized along the $a$, $b$, and $c$ directions and the corresponding images of crystals. The incident directions for $E // a$, $b$, and $c$ are $k // b$, $a$, and $a$, respectively. The data above 2.4 eV for $E // b$ and $c$ exceed detection limit due to the intense absorptions.

Relations between the color and orientation of crystal are presented in Fig. 2(a). The crystal is orange when viewed along the $a$ direction ($k // a$), while green when viewed along the $b$ or $c$ direction ($k // b, c$). The green colors for $k // b$ and $c$ are also slightly different to each other. Absorption spectra reveal the origin of these differences in color. A transmission peak at around 2.4 eV, which corresponds to the energy of green light, does not exist for $k // a$ but for $k // b$ and $c$, giving the green colors for the latter. For $k // a$, only reddish light with lower energies are transmitted, giving the orange color. There is also a weak peak in the reddish light region for $k // c$, not for $k // b$, which results in the subtle difference in the green colors between $k // b$ and $c$.

The visible color is the consequence of the summation of absorptions for a set of orthogonally polarized lights perpendicular to the propagating direction. Each component of absorption with the direction of polarization of light ($E$) parallel to $a$, $b$, or $c$ can be separated by using an incident polarizer and is shown in Fig. 2(b). The $Ca_3ReO_5Cl_2$ crystal exhibits totally different vivid colors depending on the direction of polarization of light: they are green, red, and yellow for $E // a$, $b$, and $c$, respectively. Thus, for example, the orange color for $k // a$ in Fig. 2(a) comes from the combination of the red ($E // b$) and the yellow ($E // c$) colors for polarized lights in Fig. 2(b).

The optical density [= $-\log_{10}$(Transmission)] spectra for linearly polarized lights in the energy range including visible light (1.6 ~ 3.3 eV) are shown in Fig. 2(b). The spectrum for $E // a$ has two absorption peaks centered at 1.4 and 1.9 eV as well as another peak above 3.0 eV. The corresponding two peaks also exist for $E // c$, while the peak at ~ 1.9 eV is missing for $E // b$. In addition, commonly for $E // b$ and $c$, there is a very strong absorption above 2.4 eV. For $E // b$, another absorption peak exists at ~ 2.3 eV, just below the continuous absorption above 2.4 eV, as discussed below.

Here we would like to consider the origin of the observed absorptions based on the strength and energy of them. In general, there are two relevant processes for light absorptions in transition metal compounds: one is the $d$-$d$ transition in a transition metal ion and the other is the charge transfer (CT) transition between ligands and a transition metal ion. The former is forbidden by the Laporte rule for centrosymmetric coordinations and allowed for non-centrosymmetric coordinations, as mentioned before, while the latter is typically Laporte-allowed and can be stronger than the former. Moreover, the $d$-$d$ transition causes narrower absorption peaks than the CT transition, because the $d$ levels are usually sharper than the $p$ band of ligands. Furthermore, CT peaks often lie at higher energy than $d$-$d$ peaks. Based on these general characteristics, it is reasonable to assign the three pronounced absorption peaks at 1.4, 1.9, and 2.3 eV to intra-atomic $d$-$d$ transitions at the $Re^{6+}$ ion and also to assign the strong and continuous absorptions above 3.0 eV for $E // a$ and above 2.4 eV for $E // b$ and $c$ to CT transitions between the oxygen $2p/3s$ and Re $5d$ levels. The pleochroism of $Ca_3ReO_5Cl_2$ should arise from the strong polarization dependence of the three $d$-$d$ transitions and the CT transitions in the visible light region.

The CFSs of the $5d$ orbitals in $Ca_3ReO_5Cl_2$ are intuitively predicted as illustrated in Fig. 3. In the regular octahedral coordination ($O_h$), the $d$ orbitals split into doubly degenerate $e_g$ orbitals pointing toward the ligands at higher energy and triply degenerate $t_{2g}$ orbitals aside from the ligands at lower energy. In the square pyramid ($C_{4v}$) with one of the apical ligands removed, the degeneracies of the $e_g$ and $t_{2g}$ orbitals are further lifted: the $d_{z^2}$ and $d_{yz}/d_{zx}$ orbitals have lower energies owing to reduced Coulomb repulsions from the apical ligand. For $Ca_3ReO_5Cl_2$, in addition, the basal plane of the $ReO_5$ SP is slightly distorted from a regular square to a trapezoid with keeping the mirror plane perpendicular to the $b$ axis. Thus, the "true" symmetry is, $C_s$, in which the degenerate $d_{yz}/d_{zx}$ orbitals should split into two linear combinations of $d_{yz}$ and $d_{zx}$: the $d_{xz-yz}$ orbital extended in the $ab$ plane and the $d_{xz+yz}$ orbital in the $bc$ plane. The energy splitting of these orbitals may be relatively small because the trapezoidal distortion of the basal plane (= 1 − distance$_{O1-O1}$ / distance $_{O2-O2}$) is as small as 2.4%.

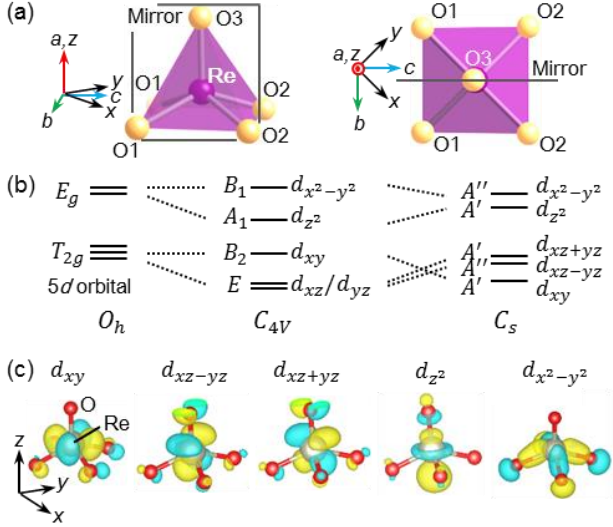

**Figure 3.** (a) Local coordination of Re in the ReO$_5$ square pyramid. One ReO$_5$ SP with the local $z$ axis along the crystallographic $a$ axis and with the $x$ and $y$ axes toward the planer oxygen atoms along the crystallographic [011] and [0$\bar{1}$1] directions, respectively, is considered. (b) Schematic representations of the evolution of the crystal field splitting of the 5$d$ levels in Ca$_3$ReO$_5$Cl$_2$. The degeneracy is lifted with reducing the symmetry from the octahedron ($O_h$) to square pyramid ($C_{4v}$) and further to the actual symmetry of Ca$_3$ReO$_5$Cl$_2$ ($C_s$) with only a mirror plane perpendicular to the $b$ axis remaining. The characters left and right of each horizontal bar indicate the corresponding representations of the group and the orbitals, respectively. An experimentally determined scheme of the crystal field splitting is shown for the $C_s$ symmetry. (c) Wannier orbitals corresponding to the $d_{xy}$, $d_{xz-yz}$, $d_{xz+yz}$, $d_{z^2}$, and $d_{x^2-y^2}$ orbitals from the DFT calculations.

Another important fact to be taken into account is that the Re ion shifts from the basal plane toward the apical oxygen. The distances between the Re and oxide atoms are much shorter for the apical oxygen (1.716 Å for Re–O3) than those for the basal oxygens (1.899 Å for Re–O1 and 1.920 Å for Re–O2). This means that increased Coulomb repulsions from the apical oxygen push up the $d_{z^2}$ and $d_{xz-yz}/d_{xz+yz}$ levels and push down the $d_{x^2-y^2}$ and $d_{xy}$ levels. As the result, the lowest occupied state may be either of $d_{xy}$, $d_{xz-yz}$, or $d_{xz+yz}$. The final energy diagram of the 5$d$ orbitals of Ca$_3$ReO$_5$Cl$_2$ is not trivial and should be determined experimentally.

In order to capture the electronic structure of Ca$_3$ReO$_5$Cl$_2$, we performed first-principles calculations based on the DFT. It is found that Re 5$d$ and O 2$p$ states are strongly hybridized and form bonding states at around −9 eV and anti-bonding states between 0 and 2.5 eV with respect to the Fermi energy ($E_F$); between them, non-bonding states of chlorine and oxygen have dominant contributions (Supplementary). The anti-bonding 5$d$/2$p$ states are almost non-dispersive and can be projected into localized Wannier orbitals, as depicted in Fig. 3(c). They nicely correspond to the bare 5$d$ orbitals discussed above. Thus calculated energy diagram consists of the $d_{xy}$, $d_{xz-yz}$, $d_{xz+yz}$, $d_{z^2}$, and $d_{x^2-y^2}$ orbitals in order from low energy, as depicted for the $C_s$ symmetry in Fig. 3(b).

In order to get experimental information about the energy diagram of the 5$d$ orbitals, we carried out XAS experiments. The $L_{III}$ edge of Re, which corresponds an excitation from the Re 2$p$ state to the lowest excited 5$d$ state, were examined. The lowest excited 5$d$ state must be the lowest unoccupied state, because an excitation to the lowest occupied 5$d$ state is unlikely as it costs a Coulomb repulsion from the occupied electron. As shown in Fig. 4, the XAS spectra for x-rays polarized parallel to the $a$ and $b$ axes are almost identical below 10.535 keV, while the absorption edge for $c$-polarized x-rays is located approximately 1 eV higher in energy. This result indicates that the $c$ axis is unique for the lowest unoccupied orbital. Among the 5 orbitals illustrated in Fig. 3(c) and also in Fig. 5, only the $d_{xz-yz}$ orbital extending in the $ab$ plane satisfies this condition and is identified as the lowest unoccupied state. The result is consistent with the first-principles calculations shown in Fig. 3(b), and strongly suggests that the $d_{xy}$ orbital is the highest occupied state.

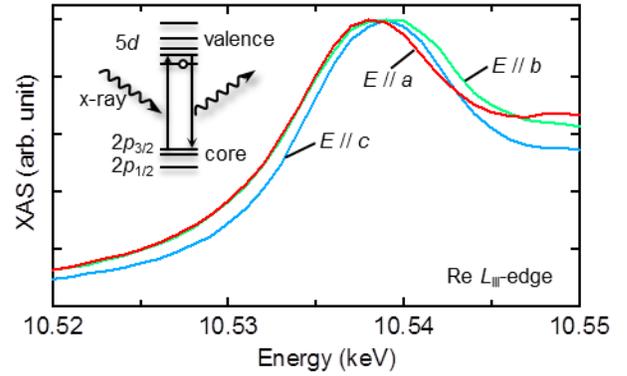

**Figure 4.** X-ray absorption spectra measured at the Re $L_{III}$ edge for incident x-rays polarized along the $a$, $b$, and $c$ directions. The inset illustrates schematic diagram of the x-ray absorption and emission processes at the Re $L_{III}$ edge.

Now, we would like to explain the unique optical property of Ca$_3$ReO$_5$Cl$_2$ based on the energy diagram and the optical selection rule of electronic transitions. According to Fermi's golden rule, the probability of an electronic transition, namely the intensity of absorption, is proportional to the transition dipolar moment as,

$I \propto |\langle g|P|e\rangle|^2$,

where $e$, $g$, and $P$ represent the excited state, the ground state, and the dipolar moment of light, respectively. Transitions are allowed when the transition dipolar moment is finite, which is realized when the total combination of the parities of the three components is even. For Ca$_3$ReO$_5$Cl$_2$, since the local symmetry at the Re site is .m. in $C_s$ with a mirror plane perpendicular to the $b$ axis, the parity of an orbital with respect to this mirror plane is to be considered. As shown in Fig. 5, the parities of the $d$ orbitals are even ($d_{xy}$), odd ($d_{xz-yz}$), even ($d_{xz+yz}$), even ($d_{z^2}$), and odd ($d_{x^2-y^2}$).

Lights polarized along the three crystallographic axes can induce different electronic transitions. Provided that the ground state is the $d_{xy}$ orbital with even parity from the DFT calculations and the XAS results, an electronic transition is allowed only when the excited state and the polarization of light have the same parity. Thus, a light polarized along the $b$ axis, which has odd parity with respect to the mirror plane, induces excitations to the odd parity orbitals, $d_{xz-yz}$ and $d_{x^2-y^2}$, as shown in Fig. 5. The absorption peaks observed at 1.4 and 2.3 eV for $E // b$ must correspond to excitations to these orbitals, respectively. On the other hand, a light polarized along the $a$ or $c$ axis has even parity with respect to the mirror plane and allows excitations to the $d_{xz+yz}$ and $d_{z^2}$ orbitals with even parity. Thus, the absorption peaks observed at 1.4 and 1.9 eV for $E // a$ and $c$ are assigned to excitations to these orbitals, respectively. The fact that the excitations to $d_{xz-yz}$ for $E // b$ and to $d_{xz+yz}$ for $E // a$ and $c$ are observed at almost the same energy of 1.4 eV means that the energy splitting of $d_{xz-yz}$ and $d_{xz+yz}$ is relatively small, as expected from the small distortion of the basal plane of the ReO$_5$ SP. Therefore, the absorption at 1.4, 1.9, and 2.3 eV are reasonably assigned to the excitations from $d_{xy}$ to nearly degenerated $d_{xz-yz}/d_{xz+yz}$, $d_{z^2}$, and $d_{x^2-y^2}$, respectively. Hence, we are successful in explaining the observed optical properties based on the CFSs of the Re 5$d$ orbitals.

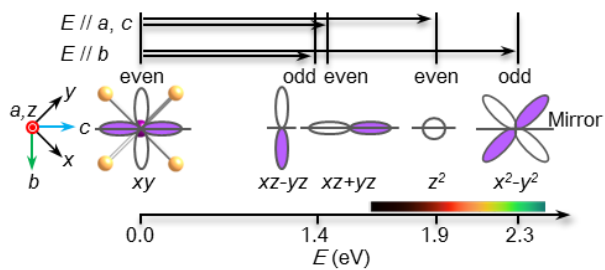

**Figure 5.** Schematic energy diagram of the Re 5$d$ levels and allowed optical transitions in Ca$_3$ReO$_5$Cl$_2$. The corresponding $d$ orbitals are schematically depicted below.

It is noted that the CT peak appears above 3.0 eV for $E // a$ and above 2.4 eV for $E // b$ and $c$. This difference may come from different CT paths from the oxygen 2$p$ to the Re 5$d$ levels. Since all the ReO$_5$ SPs align along the $a$ direction, an $a$-polarized light can induce a CT between the $p_z$ orbital of the apical oxygen and the $d_{z^2}$ orbital of the Re. On the other hand, a $b$($c$)-polarized light induces a CT between the basal oxygen and the other Re 5$d$ orbitals ($d_{xy}$, $d_{xz-yz}$, $d_{xz+yz}$, and $d_{x^2-y^2}$). Possibly, a strong hybridization between the $p_z$ and $d_{z^2}$ orbitals enlarges the CT energy for $E // a$.

Two characteristics of Ca$_3$ReO$_5$Cl$_2$ play important roles in the emergence of the distinct pleochroism. One is the unique SP coordination of Re, and the other is the large and complex CFSs of the 5$d$ orbitals compared with those of typical 3$d$ orbitals. The lack of inversion symmetry in the SP coordination allows intense $d$-$d$ transitions, and the low symmetry ($C_s$) lifts all the degeneracy of 5$d$ orbitals, which gives rise to multiple optical excitations depending on the direction of the light polarization according to the optical selection rule. A similar SP coordination is found in 3$d$ transition metal oxides such as vanadium oxides. However, most vanadates do not exhibit pleochroism. This is because typical CFSs in the SP coordination for 3$d$ orbitals are as large as 1.5 eV, which is lower than the energy of visible light[21]. In contrast, the CFS of spatially extended 5$d$ orbitals can be larger owing to stronger Coulomb repulsion exerted from the ligands. This is in fact the case for Ca$_3$ReO$_5$Cl$_2$. As a result of these two features, the optical selection rule of electronic transition manifests itself in the pleochroism of Ca$_3$ReO$_5$Cl$_2$. Note that such a vivid color change as observed in Ca$_3$ReO$_5$Cl$_2$ is accidentally achieved and may be unique in nature.

In addition to the remarkable optical property, Ca$_3$ReO$_5$Cl$_2$ exhibits an interesting magnetism arising from the 5$d^1$ electron. This compound is a Mott insulator and a quantum magnet carrying spin 1/2. Surprisingly, the magnetic property is completely one-dimensional in spite of the three-dimensional crystal structure. This is probably due to the specific arrangement of the occupied $d_{xy}$ orbitals, which gives rise to highly anisotropic magnetic interactions, to cause an embedded one dimensionality. The detail of magnetism of this compound will be reported elsewhere. Moreover, interplay between the optical property and the magnetism would be explored in the future study; one would expect a dramatic change in magnetism with optical excitations.

### CONCLUSIONS

In summary, we report on a new 5$d$ transition metal oxychloride Ca$_3$ReO$_5$Cl$_2$ exhibiting a strong and beautiful pleochroism. This unique optical property is reasonably explained based on the selection rule of electronic transitions between 5$d$ orbitals of rhenium. The key to realize this property is that the Re$^{6+}$ ion has an unusual square-pyramidal coordination, which lifts the orbital degeneracy to gain the complex crystal field splitting in the visible light energy. Our finding demonstrates that many interesting compounds still remain unexplored in 5$d$ transition metal compounds compared with 3$d$ compounds.

### ASSOCIATED CONTENT

**Supporting Information**

The Supporting Information is available free of charge on the ACS Publications website at DOI: 10.1021/jacs.??????.
Crystallographic data (CIF)
Calculated electronic structure and crystallographic data (PDF)
Movie of color change (AVI)

### AUTHOR INFORMATION

**Corresponding Authors**

*dhirai@issp.u-tokyo.ac.jp
*hiroi@issp.u-tokyo.ac.jp

**ORCID**

Daigorou Hirai: 0000-0003-2883-4376

**Notes**
The authors declare no competing financial interest.

### ACKNOWLEDGMENT S

The authors are grateful to N. Kojima for insightful discussions and T. Nakajima, V. Ukleev, Y. Yamasaki, and H. Ohsumi for their technical support. The Synchrotron XAS experiments were performed at SPring-8 with the approval of RIKEN (Proposal No. 20160093). This work was financially supported by JSPS KAKENHI Grant Number JP15K17695, the Core-to-Core Program for Advanced Research Networks given by Japan Society for the Promotion of Science, OPERANDO-OIL, JST-SENTAN, APSA, and NEDO in Japan.